\documentclass[10pt, onecolumn]{IEEEtran}
\usepackage{caption}

\usepackage{graphicx}
\graphicspath{ {Images/} }
\ifCLASSINFOpdf

\else
 \fi

\begin{document}
\pagenumbering{gobble}
\title{Artificial Intelligence as an Enabler for Cognitive
Self-Organizing Future Networks}

\author{\IEEEauthorblockN{Siddique Latif$^1$, Farrukh Pervez$^1$, Muhammad Usama$^2$, Junaid Qadir$^2$\\}
\IEEEauthorblockA{$^1$National University of Science and Technology, Islamabad, $^2$Information Technology University (ITU),
Lahore}}


%


\maketitle

%


%
The explosive increase in number of smart devices hosting sophisticated applications is rapidly affecting the landscape of
information communication technology industry. Mobile subscriptions, expected to reach 8.9 billion by 2022
\cite{cerwall2016ericsson}, would drastically increase the demand of extra capacity with aggregate throughput anticipated to be
enhanced by a factor of 1000 \cite{andrews2014will}. In an already crowded radio spectrum, it becomes increasingly difficult to
meet ever growing application demands of wireless bandwidth. It has been shown that the allocated spectrum is seldom utilized by
the primary users and hence contains spectrum holes that may be exploited by the unlicensed users for their communication. As we
enter the Internet Of Things (IoT) era in which appliances of common use will become smart digital devices with rigid performance
requirements (such as low latency, energy efficiency, etc.), current networks face the vexing problem of how to create sufficient
capacity for such applications. The fifth generation of cellular networks (5G) envisioned to address these challenges are thus
required to incorporate cognition and intelligence to resolve the aforementioned issues. Cognitive radios (CRs) and
self-organizing wireless networks are two major technologies that are envisaged to meet the future needs of such next generation
wireless networks.

CRs are intelligent and fully programmable radios that can dynamically adapt according to their prevalent environment. In other
words, they sense the spectrum and dynamically select the clearer frequency bands for better communication in the most prevailing
conditions. In this way, CRs can adaptively tune their internal parameters to optimize the spectrum usage, transmission waveform,
channel access methods and modulation schemes with enhanced coverage. However, it is due to the recent advancements in machine
learning, software defined radio (SDR) that CR is able to emerge from simulation environment to the real-time applications
\cite{qadir2016artificial}.

The overwhelming traffic growth coupled with the greedy approach towards high quality of service (QoS) has been a major challenge
for current wireless systems in terms of network resources and QoS. A new paradigm for wireless communication called 5G has been
envisioned to address these challenges. The major component of the envisioned 5G scheme is Self-Organizing Network (SON). SON is
a relatively new concept in perspective of wireless cellular networks, it refers to an intelligent network that learns from its
immediate environment, while autonomously adapting accordingly to ensure reliable communication. In fact, SON underlines new
aspect for automation of future networks in 5G era.

 The sensing, learning and reasoning behavior of both CRs and SON is achieved by extensively using artificial intelligence (AI)
and machine-learning techniques. The CRs are an evolved form of SDRs, realized by the embodiment of cognitive engine (CE) that
exploits the AI techniques for the cognitive behavior to decide optimally.

\begin{figure}[!ht]
\centering
\captionsetup{justification=centering}
\includegraphics[width=5.2in]{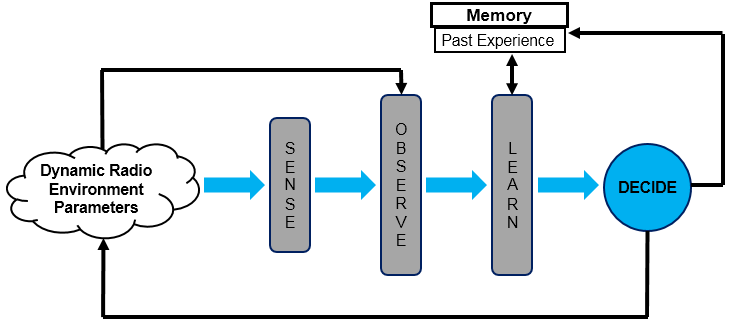}
\caption{Learning process in cognitive radios}
\label{1}
\end{figure}

\begin{figure}[!hb]
\centering
\captionsetup{justification=centering}
\includegraphics[width=5.2in]{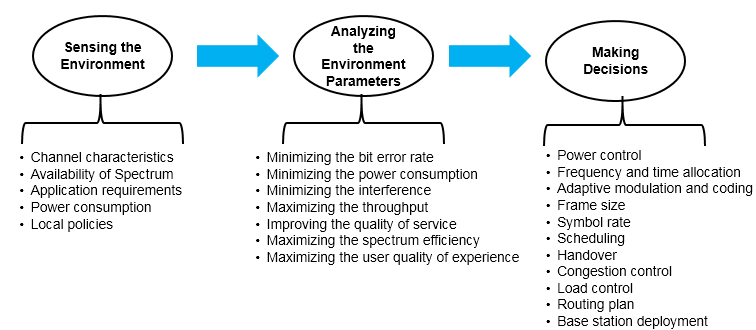}
\caption{The cognitive cycle of CR}
\label{2}
\end{figure}

 The CR network (CRN) follows the cognitive cycle, for unparalleled resource management and better network performance. Cognitive
cycle, as illustrated in figure \ref{1}, begins with sensing of dynamic radio environment parameters, subsequently observing and
learning recursively the sensed values for reconfiguration of the critical parameters in order to achieve the desired objectives.
Cognitive cycle is elaborated in figure \ref{2}, which highlights the parameters that CR needs to quantify in order to utilize
the available spectrum without affecting primary user's performance. The sensed parameters are treated as stimuli for achieving
different performance objectives, for instance, minimizing the bit error rate or minimizing the power consumption etc.
\cite{abbas2015recent}. To achieve the aforementioned objectives, CR adaptively learns deciding optimal values for various
significant variables such as power control, frequency band allocation, etc. \cite{abbas2015recent}.

CR incorporates machine learning techniques for dynamic spectrum access (DSA) and capacity maximization. AI-based techniques for
decision making such as optimization theory, Markov decision processes (MDPs), and game theory is used to encompass a wide range
of applications \cite{qadir2016artificial}. The popular learning techniques used in cognitive cycle are support vector machine
(SVM), artificial neural networks (ANNs), metaheuristic algorithms, fuzzy logic, genetic algorithms, hidden Markov models (HMMs),
Bayesian learning, reinforcement learning, multi-agent systems. Fuzzy logic theory has been used for effective bandwidth,
resource allocation, interference and power management
\cite{qadir2016artificial,kaur2010fuzzy,aryal2012novel,matinmikko2013fuzzy}. Genetic algorithms (GAs) have been employed for CRs
spectrum and parameters optimization \cite{qin2012artificial,chen2010genetic,moghal2010spectrum}. ANNs have been incorporated to
improve the spectrum sensing and adaptively learn complex environments, without substantial overhead
\cite{tan2013frequency,zhang2012cooperative}. Game theory enables CRNs to learn from its history, scrutinize the performance of
other CRNs, and adjust their own behavior accordingly \cite{han2012game,bellhouse2007problem}. In multi-agent domain,
reinforcement learning (RL) a reward-penalty based technique, which reinforces immediate rewards to maximize long term goals has
been employed for efficient spectrum utilization \cite{barve2012dynamic}, minimum power consumption \cite{zhou2012reinforcement}
and filling the spectrum holes dynamically \cite{arunthavanathan2013reinforcement}. SVM, a supervised classification model, is
being utilized for channel selection \cite{huang2009design}, adaptation of transmission parameters \cite{hu2014mac} and
beam-forming design \cite{lin2013bf}. In CRNs, HMMs have been widely used to identify spectrum holes detection
\cite{mukherjee2014spectrum}, spectrum handoff \cite{pham2014spectrum}, and competitive spectrum access \cite{li2014markov}. The
range of AI-based techniques are not limited to the above mentioned applications, other applications of AI in CRNs are expressed
in \cite{qadir2016artificial,abbas2015recent}. By combining increasing spectrum agility, context aware adaptability of CR and AI
techniques, CR has become an increasingly important feature of wireless systems. IEEE 802.16h has recommended CR as one of its
key features and a lot of efforts are being made to introduce CR features in 3GPP LTE-Advance.

\begin{figure}[!ht]
\centering
\captionsetup{justification=centering}
\includegraphics[width=4.5in,height=3in]{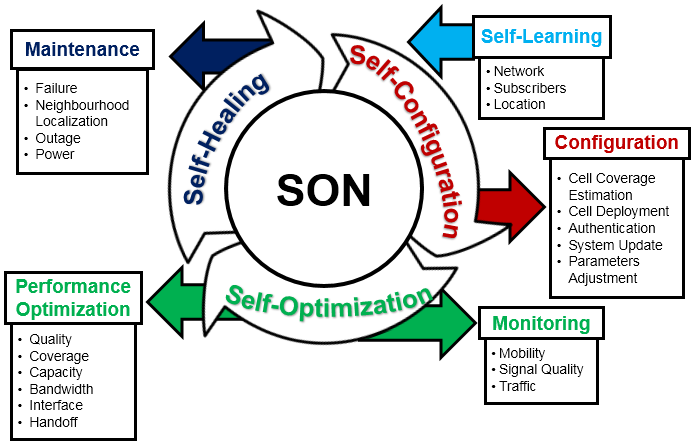}
\caption{Illustration of AI-based self-organization in the networks}
\label{3}
\end{figure}

The rapid proliferation of multi-radio access technology-disparate smart devices has resulted in complicated heterogeneous mobile
networks thus making configuration, management and maintenance cumbersome and error-prone. 5G, expected to handle diverse devices
at a massive scale, is foreseen as one of the most complicated networks and hence extensive efforts are being carried out for its
standardization. In the recent years, SONs, as depicted in figure \ref{3}, have gained significant attention regarding
self-configuration, self-optimization and self-healing of such complex networks. The idea behind SONs is to automate network
planning, configuration and optimization jointly in a single process in order to minimize human involvement. The planning phase,
which includes ascertaining cells locations, inter-cell connecting links and other associated network devices as well as
parameters, precedes the configuration phase \cite{wang2015artificial}. Self-configuration means that a newly deployed cell is
able to automatically configure, test and authenticate itself and adjust its parameters such as transmission power, inter-cell
interference etc. in a plug and play fashion \cite{wang2015artificial}. Self-healing allows trouble-free maintenance and enables
networks to recover from failures in an autonomous fashion. In addition, it also helps in routine upgrades of the equipments in
order to remove legacy bugs. Self-optimization is the ability of the network to keep improving its performance with respect to
various aspects including link quality, coverage, mobility and handoff with an objective to achieve network level goals
\cite{wang2015artificial}. Since AI-based techniques are capable to handle complex problems in large systems intrinsically, these
techniques are now being proposed to achieve Self Organization (SO) in 5G.

\begin{figure}[!ht]
\centering
\captionsetup{justification=centering}
\includegraphics[width=4.5in,height=3.2in]{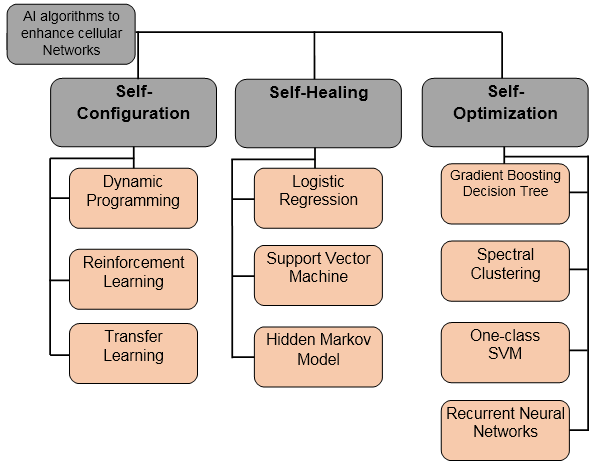}
\caption{AI algorithms for 5G}
\label{4}
\end{figure}

Self-configuration, in cellular networks, refers to the automatic configuration of initial parameters---neighbouring cells list,
IP Addresses and radio access parameters---by a node itself. AI techniques like Dynamic Programming (DP), RL and Transfer
Learning (TL) may be employed in 5G to automatically configure a series of parameters to render best services. RL, as opposed to
DP which initially builds the environment model to operate, is a model free learning technique that iterates through to reach
optimal strategy and may yield superior results in dynamically changing radio conditions \cite{liintelligent}. Self-healing is
about automatic fault detection, its classification and initiating necessary actions for recovery. Irregularities and anomalies
in network may be timely spotted to further restore the system by leveraging different AI based sensing techniques like Logistic
Regression (LR), SVM and HMM \cite{liintelligent}. Self-optimization includes continuous optimization of parameters to achieve
system-level objectives such as load balancing, coverage extension, and interference avoidance. AI techniques that may be
exploited to optimize provisioning of QoS to various services mainly belong to the class of unsupervised learning. Besides
Gradient Boosting Decision Tree (a supervised learning technique), Spectral Clustering, One-class SVM and Recurrent Neural
Networks are few examples in this regard \cite{liintelligent}. Figure \ref{4} summarizes the AI algorithms that can be utilized
to enhance cellular networks performance.

AI techniques may also exploit network traffic patterns to predict future events and help pre-allocate network resources to avoid
network overloading. Furthermore, user-centric QoS-provisioning across tiers of heterogeneous cells may also be granted using AI
\cite{liintelligent}. Similarly, GAs are employed for cell planning and optimization of coverage with power adjustment
\cite{ho2009evolving}. GAs are also suited for the problem of finding the shortest path routing in a large scale dynamic networks
\cite{mehboob2016genetic}. Wenjing et al. in \cite{wenjing2012centralized} proposed an autonomic particle swarm compensation
algorithm for cell outage compensation. The study in \cite{fan2014self} introduces the self-optimization technique for the
transmission power and antenna configuration by exploiting the fuzzy neural network optimization method. It integrates fuzzy
neural network with cooperative reinforcement learning to jointly optimize coverage and capacity by intelligently adjusting power
and antenna tilt settings \cite{fan2014self}. It adopts a hybrid approach in which cells individually optimize respective radio
frequency parameters through reinforcement learning in a distributed manner, while a central entity manages to cooperate amongst
individual cells by sharing their optimization experience on a network level \cite{fan2014self}. Cells iteratively learn to
achieve a trade-off between coverage and capacity through optimization, since increase in coverage leads to reduction in
capacity, while additionally improving energy efficiency of the network \cite{fan2014self}. ANNs can also be effectively utilized
for the estimation of link quality \cite{caleffi2009bio}. Mobile devices in an indoor environment have also been localized
through the use of ANNs \cite{ahad2016neural}. In fact, AI-based techniques enable network entities to automatically configure
their initial parameters before becoming operational \cite{wang2015artificial}, adaptively learn radio environment parameters to
provide optimal services \cite{liintelligent}, autonomously perform routine maintenance and upgrades and recover from network
failures \cite{wang2015artificial,liintelligent}.

In view of the continued proliferation of smart devices, we anticipate that CRs and SON will soon become the basic building
blocks of future wireless networks. These technologies will transform future networks into an intelligent network that would
encompass user preferences alongside network priorities/constraints. AI, being the basis of both these technologies, will
continue to drive ongoing 5G standardization efforts and therefore be the cause of a major paradigm shift. AI techniques will
continue to intervene future networks finding their usage in from radio resource management to management and orchestration of
networks. In fact, we anticipate that future wireless networks would be completely dominated by AI.

%
%





%
%

\bibliographystyle{IEEEtran}
\bibliography{Blog_AI}

\end{document}